\documentclass[]{proarticle}
\usepackage{multicol}
\usepackage{graphicx}
\usepackage{amssymb}
\usepackage{epsfig}

\def\bea{\begin{eqnarray}}
\def\eea{\end{eqnarray}}

\def\d{\partial}

\def\m{\mu}
\def\n{\nu}

\def\t{\tau}

\def\~{\widetilde}

\def\bY3{\bar Y_{,3}}
\def\Y3{Y_{,3}}

\def\Y{{\bar Y}}

\def\`{\dot}
\def\be{\begin{equation}}
\def\ee{\end{equation}}
\def\bea{\begin{eqnarray}}
\def\eea{\end{eqnarray}}

\def\fn{\footnote}

\def\mn{{\mu\nu}}

\begin{document}
\title{Gravitating bag as a coherent system of the point-like and dressed electron}
\author{A. Burinskii}
\maketitle
\address{
Gravity Research Group, NSI Russian Academy of Sciences,
 B.Tulskaya 52, 115191 Moscow, Russia.}
\eads{bur@ibrae.ac.ru}
\begin{abstract}Gravitational and electromagnetic fields of an electron correspond
to over-rotating Kerr-Newman (KN) solution, which has a naked singular ring and two-sheeted topology. This solution is regularized by a solitonic source, in which singular interior is replaced by a vacuum bubble filled by the Higgs field in a false-vacuum state.
Field model of this KN bubble has much in common with the famous MIT and SLAC bag models, but the geometry is ``dual'' (turned inside out), leading to consistency of the KN bag model  with gravity. Similar to other bag models, the KN bag is compliant to deformations, and under rotations it takes an oblate ellipsoidal form, creating a circular string along the border. Electromagnetic excitations of the KN bag generate stringy traveling waves which deform the bag, creating a traveling singular pole, included in a general bag-string-quark complex.  The dressed electron may be considered in this model as a coherent excitation of this system, confining  the point-like electron (as a quark) in state of zitterbewegung.\end{abstract}



\begin{multicols}{2}
\section{Introduction}
The Kerr-Newman rotating black hole solution has gyromagnetic ratio $g=2$ as that of the Dirac electron. The measurable parameters of the electron:
spin, mass, charge and magnetic moment determine the
 gravitational and electromagnetic field of electron as field of the over-rotating Kerr-Newman (KN) solution. The
corresponding space-time has topological defect -- the naked Kerr
singular ring, which forms a branch line of the Kerr space
 into \emph{two sheets:} the sheet of advanced and sheet of the retarded
fields.
\begin{figure*}[hbp]\center
\includegraphics[width=2in]{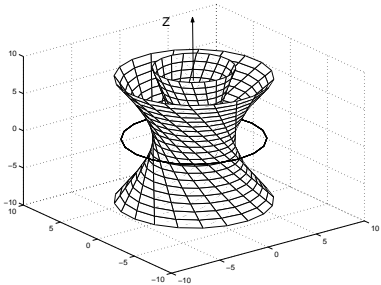}
\caption{Kerr's principal congruence of the null lines (twistors) is focused on the Kerr singular ring, forming a branch line of the Kerr space into two sheets and generating two different metrics on the same background.}
\label{fig1}
\end{figure*}

The Kerr-Schild form of metric \cite{DKS}\be g_\mn =\eta_\mn + 2H
k_\m k_\n ,\label{KSH} \ee in which $ \eta_\mn $ is metric of
auxiliary Minkowski space $M^4 ,$ and $ k_\m $ is a null vector
field, $ k_\m k^\m =0 ,$ forming the Principal Null Congruence
(PNC) $\cal K .$\fn{We use signature $(- + + +)$.} The retarded
and advanced sheets are related by a smooth transfer of the Kerr PNC
via disk $ r=0 $ spanned by the Kerr singular ring $ r=0, \
\cos\theta=0 $ (see fig.1), where $r$ is the Kerr ellipsoidal
radial coordinate.  The surface $r=0$ represents a disklike "door"
from negative sheet $r<0$ to positive one $r>0$. The null vector
fields $k^{\m\pm}(x)$ differs on these sheets, and  two
different null congruences ${\cal K}^\pm ,$ create two different
metrics \be g_\mn^\pm =\eta_\mn + 2H k_\m^\pm k_\n^\pm
\label{KSpm} \ee on the same Minkowski background $M^4.$
  Twosheetedness of the Kerr geometry caused
  search for different models of the source of KN solution,
  without mystery of the negative sheet.

Singular metric conflicts with basic principles of quantum theory
which is settled on the flat space-time and negligible
gravitation. Resolution of this conflict requires "regularization"
of space-time, which has to be done \emph{before quantization,
i.e. on the classical level}. Singular region has to be excised
and replaced by a regular core with a flat internal metric
$\eta_\mn ,$ matching with external KN solution. Long-term search
for the models of regular source (H. Keres (1966), W. Israel
(I970), V. Hamity (1976), C. L\'opez (1984) at al. See refs in
\cite{BurSol1}) resulted in appearance of the gravitating
soliton model \cite{BurSol,BurSol1} which represents a domain-wall
bubble, or a bag confining the Higgs field in a superconducting false-vacuum
state.

Such a matter regulates the KN electromagnetic (EM) field
pushing it from interior of the soliton to domain wall boundary and
results in consistency of the internal metric with  flat space required by
quantum theory. The Higgs mechanism of broken symmetry approaches this
field model with the famous MIT and SLAC bag models \cite{MIT,SLAC}.
However, the typical quartic potential  \be V(|\Phi|)=g(\bar\sigma \sigma - \eta^2)^2 ,\label{phi4} \ee used
for the self-interacting Higgs field $\Phi$, in the most of the soliton and bag-like models cannot
provide consistency with external KN solution.
 Eg, in the MIT bag model, the vacuum expectation value (vev)
 of the Higgs field $\sigma =<|\Phi|> $ vanishes inside the bag, $r<R , $ and takes non-vanishing value $ \sigma = \eta ,$
 \emph{ outside the bag,} $r> R .$
Therefore, the bag forms a well or cavity in the space-time filled by the Higgs condensate in a superconducting phase, and this condensate turns the external electromagnetic fields in the short-range ones, spoiling the external KN solutions. Similar problem appears also in the SLAC bag model.

%
\section{Field model of phase transition for gravitating bag}\label{sec2}
%
To ensure the correct external gravitational and electromagnetic (EM) fields, the symmetry breaking mechanism should be turned inside out: the Higgs condensate must be localized inside the bag, $ r <R, $ leaving the unbroken external vacuum state.

Formation of the corresponding potential turns out to be a very non-trivial problem, and to realize it,   we used in \cite{BurSol} a supersymmetric scheme of phase transition with
three chiral fields $\Phi^{(i)}, \ i=1,2,3 ,$  \cite{WesBag}. One of which, say $\Phi^{(1)} ,$ has the required radial dependence, and we chose it as the Higgs field ${\cal H} ,$  setting the new notations in accord with
$({\cal H}, Z, \Sigma) \equiv (\Phi^0, \Phi^1, \Phi^2) .$
 The required potential $V(r)=\sum _i |\d_i W|^2  $ is obtained  from the superpotential  $ W(\Phi^i, \bar \Phi^i) = Z(\Sigma \bar \Sigma -\eta^2)
+ (Z+ \m) {\cal H} \bar {\cal H},\label{W} $ where $ \m$ and  $
\eta $ are real constants.
 The condition $ \d_i W =0 $ determines two vacuum states separated by a spike of the potential $V$ by $R-\delta < r <R-\delta $:

(I) external vacuum, $r>R +\delta $, $V (r) = 0 ,$ with vanishing Higgs field ${\cal H} =0 $,   and

(II) internal state of false vacuum, $r<R-\delta $, $V (r) = 0 ,$ with broken symmetry, $|{\cal H}| = \eta = const.$

(III) intermediate region of the domain wall  phase transition $R-\delta < r <R-\delta $

\emph{The boundary of  phase transition is unambiguously determined} by matching the external KN metric
 $ g_\mn =\eta_\mn + 2H k_\m k_\n , $ where \be
H=\frac {mr -e^2/2}{r^2+a^2 \cos ^2 \theta} \ee with flat internal
metric $ g_\mn =\eta_\mn .$ It fixes the boundary at $H=0 ,$ which yields
 \be R = r_e = \frac {e^2}{2m} . \label{re} \ee
 Since $r$ is the Kerr oblate coordinate,
 the bag forms an oblate disk of the radius
 $ r_c \approx a = \frac {1}{2m}$ with thickness $r_e=\frac {e^2}{2m},$
so that  $r_e/r_c = e^2 \approx 137^{-1}.$

The choose of  L\'opes's boundary for regularization of the KN source allows us to neglect gravity inside the source and in the zone of phase transition and consider there the space-time as flat, while outside the source the gravity and EM field are unbroken.

 Inside the source  we have only the part of Lagrangian which corresponds to self-interaction of the complex Higgs
field and its interaction with vector-potential of the KN electromagnetic field $A^\m $ in the flat space-time.
 The field model in zone (III) is reduced to the Nielsen-Olesen model for a vortex string in superconducting media \cite{NO}, \be
{\cal L}_{NO}= -\frac 14 F_\mn F^\mn + \frac 12 ({\cal D}_\m {\cal
H})({\cal D}^\m {\cal H})^* + V(|{\cal H}|), \label{LNO}\ee where
$ {\cal D}_\m = \nabla_\m +ie A_\m $ is a covariant derivative,
$F_\mn = A_{\n,\m} - A_{\m,\n}  ,$  and $\nabla_\m \equiv \d_\m$ is reduced to derivative
in flat space with a flat D'Alembertian $\d_\n \d^\n = \Box  .$
For interaction of the complex Higgs field \be{\cal H}(x) = |{\cal H (x)}| e^{i\chi(x)} \ee
with the Maxwell field we obtain the
following complicated systems of the nonlinear differential equations
\bea D_\n D^\n {\cal H} &=& \d_{\bar{\cal H}} V  , \label{PhiIn} \\
\Box A_\m = I_\m &=&  e |{\cal H}|^2
(\chi,_\m + e A_\m). \label{Main} \eea
 Analysis of the equation (\ref{Main}) in \cite{BurSol,BurSol1} showed two remarkable properties of the KN rotating soliton:

 \medskip

 \textbf{(A)} The vortex of the KN vector potential $A_\m$ forms a quantum
 Wilson loop placed along the border of the disk-like source, which leads to
 \emph{quantization of the angular momentum} of the soliton,

\textbf{(B)} the Higgs condensate should \emph{oscillate} inside the source with the frequency $\omega= 2m $.

The KN vector potential has the form \cite{DKS}
\be A_\m dx^\m = - Re \
[(\frac e {r+ia \cos \theta}) (dr - dt - a \sin ^2 \theta d\phi
) \label{Am}. \ee
Maximum of the potential is reached in the equatorial plane,  $\cos \theta =0 $, at the
L\'opez's boundary of the disk-like source (\ref{re}), $r_e = e^2/2m ,$ which plays the role
of a cut-off parameter,
\be A^{max}_\m dx^\m = - \frac e {r_e} (dr - dt - a d\phi
) \label{Amax}. \ee
According to (\ref{Main}) vector field is regularized, since it cannot  penetrate deeply
inside the Higgs condensate for the phase incursion of the Higgs field.

One sees, that the $\phi-$ component of the KN vector potential $A^{max}_\phi = ea/r_e $
forms a circular flow  (Wilson loop) near the source boundary. According to (\ref{Main}),
this flow does not penetrate inside the source beyond $r< r_e -\delta ,$ since it is compensated
by  gradient of the Higgs phase $\chi,_\phi .$ Integration of this relation over the closed
$\phi-$loop under condition $I_\phi =0$ leads to the result \textbf{(A)}.
Similarly, the result \textbf{(B)} follows from (\ref{Main}) under condition $I_\phi =0$ for
the time component of the vector potential $A^{max}_0  = \frac e {2 r_e} = m/e .$

\section{Fermionic sector}

The bag models give significant progress in the fermionic sector of the extended particle-like
solutions. In the SLAC bag theory, \cite{SLAC}, the Dirac equations interacting with the classical
vev of the Higgs field $\sigma$-field  take the form \be (i\gamma ^\m \d_\m - g\sigma) \psi =0
\label{Dir-sigma}, \ee where $g$ is a dimensionless coupling parameter. One sees that the Dirac
field $\psi$ acquires from the Higgs field effective mass $m =g\sigma ,$ which takes maximal value
$m=g\eta $ outside the bag  and vanishes inside the bag. The quarks are confined, getting inside
the bag a more favorable energetic position, and this is the principal idea of the confinement
mechanism.
\begin{figure*}[hbp]\center
\includegraphics[width=3in]{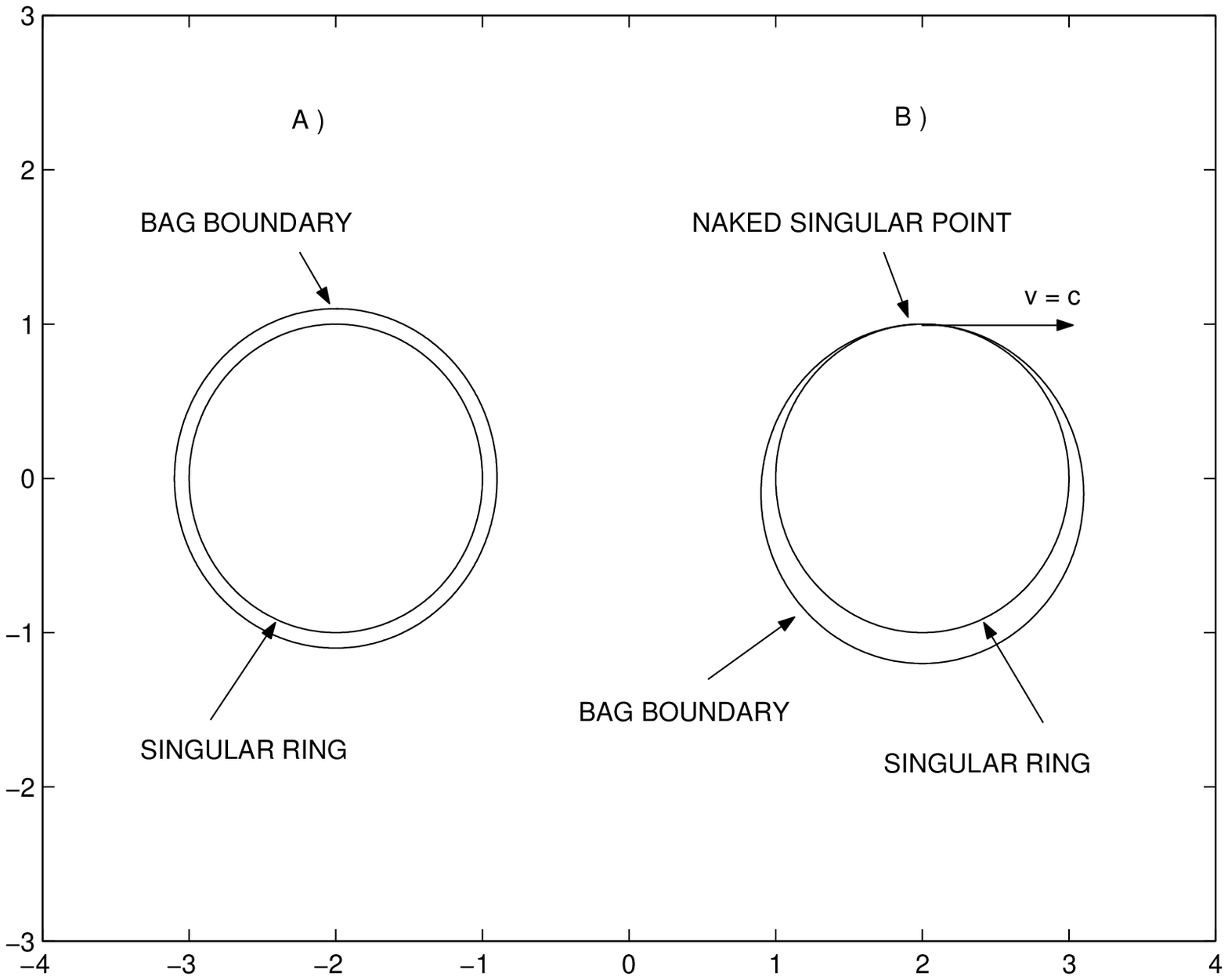}
\caption{Cut-off parameter is determined by distance $R=r_e$ from the bag boundary to singular ring in equatorial plane. A) Axially symmetric KN solution, $R= e^2/2m $.  B) Deformation of the KN bag by the lowest stringy excitation yields variable cut-off $R(\phi, \t)$ with a circulating light-like node $\phi(\t)$, creating singular pole of a bare electron.}
\label{fig2}
\end{figure*}

For gravitating bag model consistent with the KN solution, we follow this field mechanism, but must apply it ``inside out'' to have unbroken vacuum for gravitational and EM KN solution in external
region. Thus, as agreed with the external gravitational and EM field of the KN solution, the Dirac equation  will have inside the bag a nonzero bare mass $ m = g\eta $ and splits in the Weyl basis
 into the left and right chiral parts, for spinors of the Dirac bispinor
$\Psi^\dagger = (
 \phi _\alpha,
\bar\chi ^{\dot \alpha}).$
  \be
 \sigma ^\m _{\alpha \dot \alpha} i \d_\m
 \bar\chi ^{\dot \alpha}=  m \phi _\alpha , \quad
 \bar\sigma ^{\m \dot\alpha \alpha} i \d_\m
 \phi _{\alpha} =  m \bar\chi ^{\dot \alpha},
\label{Dir} .\ee  Outside the bag, $r>R-\delta ,$ the mass vanishes and these equations become independent
\be
 \sigma ^\m _{\alpha \dot \alpha} i \d_\m
 \bar\chi ^{\dot \alpha}=0, \quad
 \bar\sigma ^{\m \dot\alpha \alpha} i \d_\m
 \phi _{\alpha} = 0.
\label{Dir0} \ee
It must be consistent with spinor structure of the external KN solution which is determined by the Kerr  theorem \cite{Emerg}. The Kerr-Schild (KS) form of KN solution may be represented in  via the both Kerr congruences outgoing $k^+_\m $ or ingoing
 $k^-_\m ,$ but not via the both ones simultaneously. Taking for the \emph{physical sheet} of the KN solution the outgoing Kerr congruence $k^+_\m ,$ corresponding to the retarded EM field, we obtain the metric $g_\mn^+$, and note that the advanced fields becomes  inconsistent with this physical sheet, and being aligned with another congruence $k^-_\m ,$ they should be positioned on the separate sheet which different metric  $g_\mn^-$.
 This problem disappears inside the bag, where $H=0 $, and the space is
flat, $g^\pm = \eta_\mn ,$  and the difference between two metrics  disappears.

Due to consistency conditions the gravitational interaction drops out, and the Dirac equations take the ordinary  flat space-time form.
The consistency conditions  $k^{\m} \gamma_\m \Psi =0  $ turn into
equations for eigenfunctions of the helicity operator
 \be
(\mathbf{k}\cdot
 \mathbf{\sigma})  \phi = \phi, \quad (\mathbf{k}\cdot \mathbf{\sigma})  \chi = - \chi.
\label{align} \ee One sees that the spinors $\phi$ and
$\chi$ have opposite helicity, forming the "left-handed"  $\phi$
and "right-handed" helicity states, aligned with outgoing
direction $\mathbf{k}$ and ingoing direction $-\mathbf{k}$
correspondingly. The requirement of the consistency of the Dirac solutions with
KN solution enforces us to return the removed twosheetedness of the Kerr geometry, this time  to restore it outside the KN source. Therefore, two massless Weyl spinors $\phi_\alpha$ and $\bar\chi^{\dot\alpha} $ living on the different sheets of the external KN solution are joined at the flat space inside the bag and, obtaining mass from the Yukawa coupling, they form the Dirac bispinor.

One more novelty for the Dirac equation in the bag models is appearance of the variable mass
term,  which is determined by Higgs condensate and turns out to  be different for different regions of space-time.
The Dirac wave function, solution of the Dirac
equation with variable mass term, avoids
the region with a large bare mass, and tends to get an energetically
favorable position.
In the SLAC bag model \cite{SLAC} this problem is solved by variational approach. The corresponding Hamiltonian is
\be H(x) =\Psi^\dag (\frac 1{i} \vec \alpha \cdot \vec \nabla
+g\beta\sigma ) \Psi \label{Ham}, \ee and the energetically
favorable wave function has to be determined by minimization of
the averaged Hamiltonian ${\cal H} = \int d^3 x H(x) $
under  the normalization condition $\int d^3x \Psi^\dagger(x) \Psi(x) =1 .$
It yields
\be (\frac 1{i} \vec \alpha \cdot \vec \nabla
+g\beta\sigma ) \Psi = {\cal E} \Psi \label{H-E}, \ee
where ${\cal E}$ appears
as the Lagrangian multiplier enforcing the
normalization condition.
Similar to results of the SLAC-bag model, one expects that the Dirac
wave function will not penetrate deep in the region of large bare mass
$m =g\eta ,$ and will concentrate in a narrow transition zone at the
bag border $R-\delta <r< R+\delta $.  As it motivated in \cite{SLAC},
narrow concentration of the Dirac wave function is admissible for
scalar potential which  does not lead to the Klein paradox.
 The exact solutions of this kind are known only for
two-dimensional case, and the corresponding variational problem
 should apparently be solved numerically by
using the ansatz $\tilde \Psi =f(x)\Psi (x) ,$ in which $f(x)$ is a
variable factor for the Dirac solution based on the Weyl spinors  $\phi _\alpha,
\bar\chi ^{\dot \alpha}$ consistent  with the corresponding outgoing and ingoing Kerr congruences.

\begin{figure*}[hbp]\center
\includegraphics[width=2.5in]{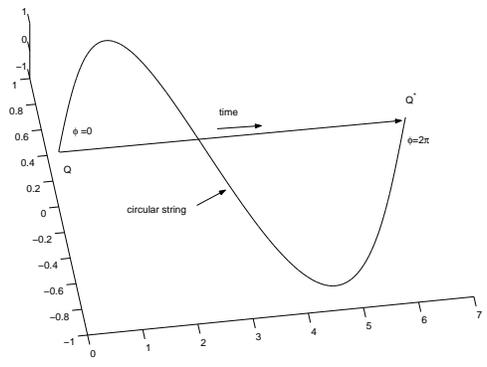}
\caption{Circular light-like string propagating along the bag border joined with time-like  zero-mode of the KN solution at  singular pole.}
\label{fig3}
\end{figure*}

\begin{figure*}[hbp]\center
\includegraphics[width=3in]{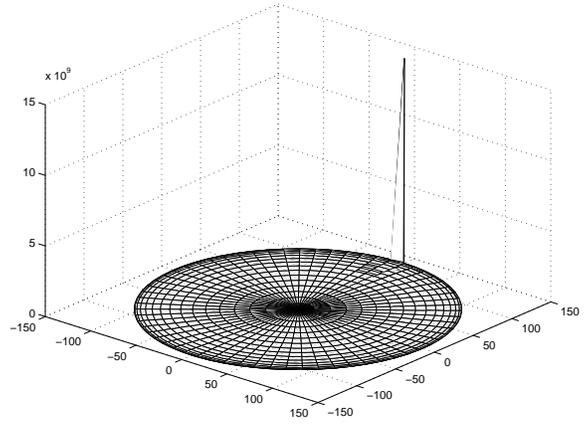}
\caption{Rotating disk-like bag -- source of the KN dressed electron and  singular pole (point-like bare electron) created by traveling wave along the string on the bag border.}
\label{fig3}
\end{figure*}

\section{Stringy deformations of bag and singular pole} Taking the bag
model conception, we should also accept the dynamical point of
view that the bags are soft and may easily be deformed \cite{SLAC}.
By deformations bags may form stringy structures. The typical
deformation of the bag models are radial and rotational excitations with
formation of the string-like  open flux-tubes. The old Dirac's model of an "extensible"
spherical electron \cite{DirBag} may also be considered as a prototype
of the bag model with spherically symmetric radial excitations.
The bag-like source of the KN solution without rotation, $a = 0 ,$  turns into the
 Dirac spherical "extensible" electron model with classical electron radius
$R=r_e=e^2/2m .$ The disk-like bag of the KN rotating
source may be considered as the Dirac spherical bag stretched by rotation to the
disk of Compton radius, $a= \hbar/2mc ,$ which corresponds to the zone of vacuum polarization of a ``dressed'' electron. The degree of oblateness of the KN
bag is approximately $ \alpha= 137^{-1}.$ In equatorial plane the boundary of the disk is very close to position of the Kerr singular ring, field structure of which agrees with the structure of fundamental string, obtained by Sen solution to low energy string theory \cite{BurSen}. Regularization of the KN source, gives to this string the cut-off parameter determined by (\ref{re}). In accord with \cite{Bur0,IvBur} this string may carry traveling waves created by the EM excitations of the KN solution, and these waves must deform the bag surface.
EM excitations of the Kerr background \cite{DKS}
 are determined by analytic function
\be A= \psi(Y,\t)/P^2 \label{A} \ee where
$ Y=e^{i\phi} \tan \frac \theta 2 $ is a complex projective angular variable,
$ \t = t -r -ia \cos \theta $ is a complex retarded-time parameter and
 $P=2^{-1/2}(1+Y\bar Y) $ for KS geometry at rest. The KS vector potential is determined by function $\psi$ as follows \cite{DKS}
\be \alpha =\alpha _\m dx^\m \\
= - Re \ [(\frac \psi {r+ia \cos \theta}) e^3 + \chi d \Y
],   \label{alpha} \ee where $\chi = 2\int (1+Y\Y)^{-2} \psi dY  \
. $

   The simplest solution $\psi=-e$ is just the KN stationary solution, which creates
 a frozen circular EM wave, which is locally plane and ``propagates'' along  the Kerr singular ring with the constant cut-off parameter $R=r_e$, see Fig.2A. It is a zero-mode excitation of the regularized Kerr string. Along with many other possible stringy waves, interesting effect shows  the simple wave solutions \be \psi = e(1 + \frac 1 Y e^{i\omega \t}) . \label{psi12}\ee
 Function $\psi$ acts on the metric through the function $H$
 \be H =\frac {mr - |\psi|^2/2} {r^2+
a^2 \cos^2\theta} \ . \label{Hpsi} \ee As we have seen in sec.2, the condition $H=0$ determines the boundary of disk $R=|\psi|^2/2m ,$ which acts as the cut-off parameter for EM field.
One sees that solution (\ref{psi12}) takes in equatorial plane $\cos \theta=0$  the form
$\psi = e(1 +  e^{-i(\phi - \omega t)}) ,$ and the cut-off parameter $R=|\psi|^2/2m =\frac {e^2}{m} (1 + \cos (\phi -\omega t)$ is not constant and vanishes at $\phi = \omega t$, creating a singular pole circulating along the Kerr singular ring, $\phi = \omega t .$

This pole may be interpreted as a point-like bare electron in the state of zitterbewegung, or as a confined quark in the conception of the bag models.

\section{Conclusion}
The mysterious problem of the source for two-sheeted Kerr geometry leads to a
gravitating soliton model, which has to retain the external KN solution. This condition cannot be realized by the usual quadratic scheme of self-interacting
Higgs field and requires a supersymmetric model of phase transition, in which the Higgs
condensate breaks the gauge symmetry only in the core of the
particle-like solution, leaving unbroken the external electromagnetic field of the Kerr-Newman solution.
 The resulting soliton model has much in common with the famous MIT and SLAC bag models. In particular, the KN bag is compliant to deformations, and takes under rotations an oblate disk-like form with a circular string placed along the border of disk. For parameters of an electron the disk acquires the Compton radius, indicating that the KN  bag  may be related to the Compton region of vacuum polarization of a dressed electron. Electromagnetic excitations of the KN bag, generated by stringy traveling waves,  deform the bag and create a bag-string-quark system which gives explanation to mystery of zitterbewegung of the Dirac electron.

\section*{Acknowledgement}
This research is supported by the RFBR grant  13-01-00602.\end{multicols}


\begin{thebibliography}{99}

\bibitem{DKS}  Debney G. C.,  Kerr R. P. and Schild A.  1969
 {\it J.\ Math.\ Phys.}  {\bf 10} 1842



\bibitem{BurSol} Burinskii  A. 2010
\emph{J. Phys. A: Math. Theor.} {\bf 43}  392001 [arXiv:
1003.2928].

\bibitem{BurSol1} Burinskii A.  2014
\emph{Int J. of Mod. Phys.} \textbf{A 29}  1450133,    [arXiv:1410.2888].

\bibitem{MIT} Chodos  A. et al. 1974
\emph{Phys. Rev.} \textbf{D 9}, 3471

\bibitem{SLAC}  Bardeen W. A. at al. 1974
\emph{Phys. Rev.} \textbf{D 11}, 1094.



\bibitem{WesBag}  Wess J.,  Bagger J.  {\it Supersymmetry and
Supergravity} (Princeton Univ. Press, Princeton, New Jersey),
1983.

\bibitem{NO}  Nielsen H. B. and  Olesen P. 1973
 \emph{ Nucl.  Phys.}  {\bf B 61},  45


\bibitem{Emerg} Burinskii A. 2014,
[arXiv:1404.5947].




\bibitem{DirBag} Dirac P.A.M.  1962
\emph{Proc. R. Soc. Lond.} \textbf{A 268},  57-67.



\bibitem{BurSen} Burinskii A.  1995
  {\it Phys.\ Rev.\  D} {\bf 52}  5826, [arXiv:hep-th/9504139].




\bibitem{Bur0}  Burinskii A.Ya. 1974
\emph{ Sov.\ Phys.\ JETP } {\bf 39} 193

\bibitem{IvBur} Ivanenko D.D. and  Burinskii A.Ya. 1975
\emph{Izv. Vuz.  Fiz.},  {\bf 5},  135


\end{thebibliography}
\end{document}